\documentclass{aa}
\usepackage{graphicx}
\usepackage{txfonts}

\begin{document}
   \title{The three-dimensional structure of sunspots}   

   \subtitle{II. The moat flow at two different heights }

   \author{H. Balthasar
          \inst{1}
          \and
          K. Muglach\inst{2}
          }

   \offprints{H. Balthasar}

   \institute{Astrophysikalisches Institut Potsdam,
              An der Sternwarte 16, D-14482 Potsdam, Germany
         \and
             Artep. Inc., Naval Research Laboratory 
             4555 Overlook Ave. SW, Washington DC 20375, USA
                          }

   \date{Received 24 July 2009; accepted 02 November 2009}

 
  \abstract
   {} 
   {Many sunspots are surrounded by a radial outflow called the moat flow.
    We investigate the moat flow at two different heights of the solar    
    atmosphere for a sunspot whose magnetic properties were reported in 
    the first paper of this series. }
   {We use two simultaneous time series taken with 
    the Transition Region And Coronal Explorer (TRACE) 
    in white light and in the UV at 170\,nm. The field-of-view
    is centered on the small sunspot NOAA\,10886 located near disk center.
    Horizontal velocities are derived by applying two different local
    correlation tracking techniques.} 
   { Outflows are found everywhere in the moat.
     In the inner moat, the velocities from the UV series are larger 
     than those from white light, whereas in the outer part of the moat we  
     find the converse result.}                                            
   { The results imply that the white light velocities 
     represent a general outflow of the quiet sun plasma in the moat,
     while UV velocities are dominated by small bright points that move
     faster than the general plasma flow. }

   \maketitle
%

\section{Introduction}

    A horizontal outflow around sunspots, called the moat flow, was first 
    detected by Sheeley (\cite{sheeley72}) while investigating the 
    Doppler shift of a magnetically insensitive spectral line (Fe\,{\sc i}\,543.4\,nm). 
    Its existence was confirmed spectroscopically by
    Balthasar et al. (\cite{FPI96}). The speed of these outflows is on  
    the order of 0.5 -- 1.0~km\,s$^{-1}$.
    
    Near-disk-center horizontal motions cannot be detected spectroscopically,
    but local correlation tracking (LCT) techniques introduced by November \& Simon  
    (\cite{lct}) allow us to follow the proper motions of intensity structures  
    such as granules or G-band bright points. Applying LCT, Rimmele (\cite{rim97})
    observed a maximum speed of 1~km\,s$^{-1}$.
    Sobotka \& Roudier (\cite{sobotka}) investigated time series of 32 sunspots
    being observed in white light (WL)  
    with the Transition Region And Coronal Explorer (TRACE),
    26 of these sunspots were observed simultaneously in the UV at 160\,nm.
    They found that moat flows occur in both young and old spots. In older spots,
    the moat is more extended, and the flow velocity is 0.410\,km\,s$^{-1}$ 
    instead of 0.380\,km\,s$^{-1}$ for young spots.
    On average, velocities are higher by 0.04 -- 0.07\,km\,s$^{-1}$ in the UV. 
    Vargas Dom\'\i nguez et al. (\cite{vargas07}, \cite {vargas08}) found moat flows 
    on the order of 0.7~km\,s$^{-1}$, but only where the adjacent penumbra has a radial
    structure. Locations without penumbra or with a non-radial penumbra have no  
    moat flow. On the other hand, Deng et al. (\cite{nadeng07}) observed a 
    persisting moat flow after the penumbra of a decaying sunspot had disappeared.

    LCT can also be applied to small magnetic areas that are found
    in the moat moving outward. 
    They are called moving magnetic features (MMFs, see Harvey \& Harvey
    \cite {harvey}) and were observed  first by Sheeley (\cite {sheeley69}).
    Sheeley discovered that these features coincide with 
    bright points in the CN-bandhead close to 388.3~nm. 
    MMFs were studied by Brickhouse \& LaBonte (\cite{brick}), who measured
    velocities of about 0.5~km\,s$^{-1}$. They claimed that the moat radius 
    is roughly twice the penumbral radius.
    Hagenaar \& Shine (\cite {Hagenaar}) found that MMFs begin at  high  
    velocities of 1.8~km\,s$^{-1}$ near the outer boundary of the penumbra,
    which is faster than average moat velocities.
    Zhang et al.~(2007) also found an outward decrease in flow
    velocity of MMFs from 0.6~km\,s$^{-1}$ to  0.35~km\,s$^{-1}$.
    Sainz Dalda \& Mart\'\i nez Pillet (\cite {alberto}) and Ravindra
    (\cite {ravindra}) found that MMFs originate in the outer penumbra.
    Intergranular G-band bright points also exhibit an
    outward motion near the spot, at least in their majority.
    Bovelet \& Wiehr (\cite {bovelet}) reported velocities on the order of 0.3~km\,s$^{-1}$
    close to the penumbral boundary.

    Kubo et al.~(\cite {kubo}) detected horizontal velocities of up
    to 1.5~km\,s$^{-1}$ by investigating the magnetic properties of MMFs.
    Choudhary \& 
    Balasubramaniam (\cite{debi}) found that there are fewer MMFs in the lower 
    chromosphere than in the photosphere. Lin et al. (\cite{lin06}) identified
    an MMF with brightenings in the chromosphere and transition region and concluded 
    that MMFs influence the dynamics in the upper layers of the solar atmosphere.
    Zuccarello et al.~(\cite {zucc}) observed MMFs streaming outward from
    a sunspot that had not developed a penumbra.
   \begin{figure*}
   \includegraphics[height=227mm]{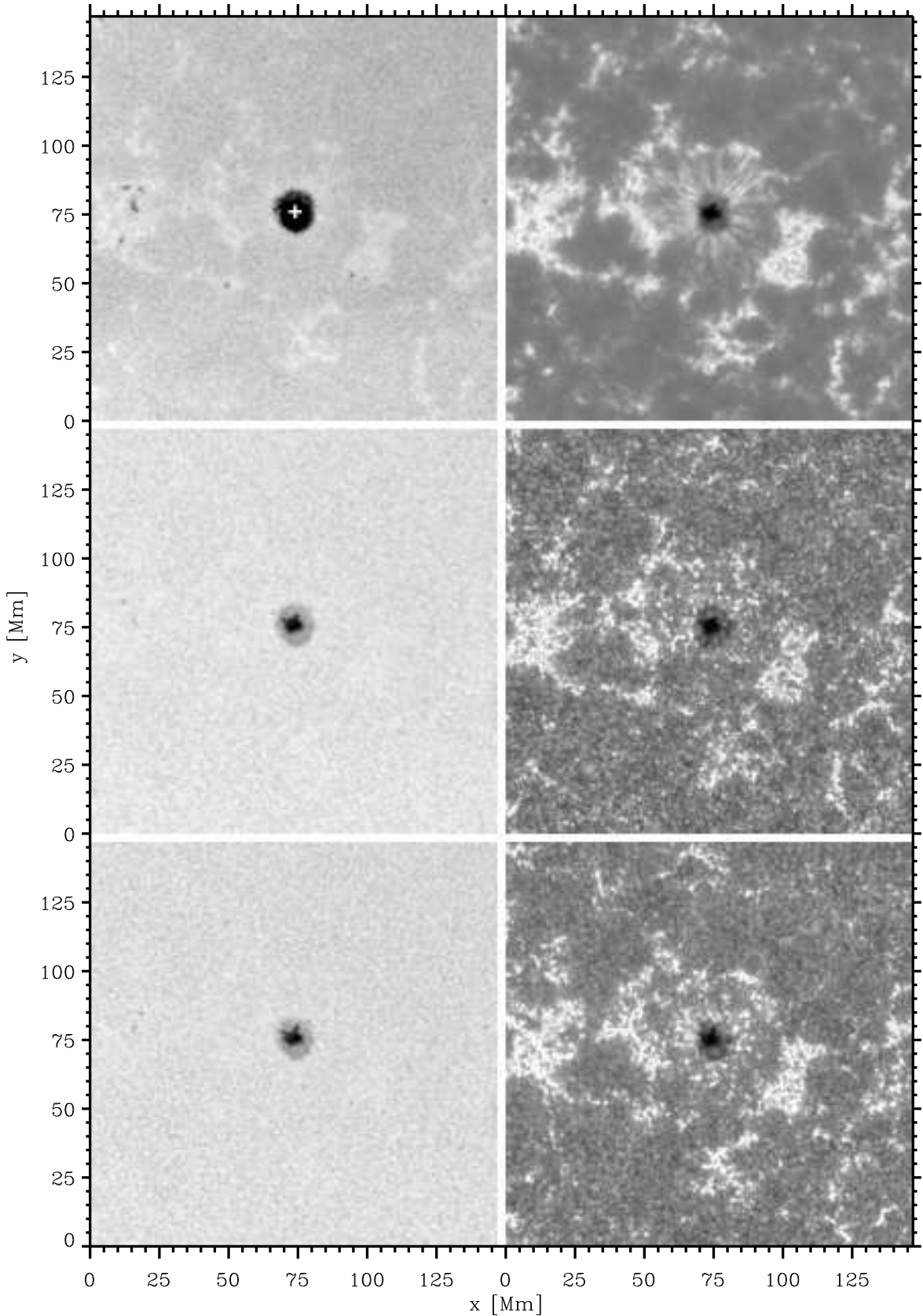}
       \caption{Images of the observed sunspot and its surroundings. 
              White light images are given in the left column and UV
              images at 170\,nm in the right one. The upper row shows 
              images averaged over the complete time series, the middle row shows
              images at the start of the series, and the bottom row those    
              at the end. The dark parts of the spot are suppressed in the 
              averaged WL image to enhance the visibility of bright network
              structures that delineate the moat. UV images are displayed   
              on a logarithmic scale. The white cross marks the center of
              the spot. The center of the solar disk is towards the
              lower edge.
            }
         \label{fig-images}
   \end{figure*}
%
%
   \begin{figure*}
   \centering
   \includegraphics[width=12cm]{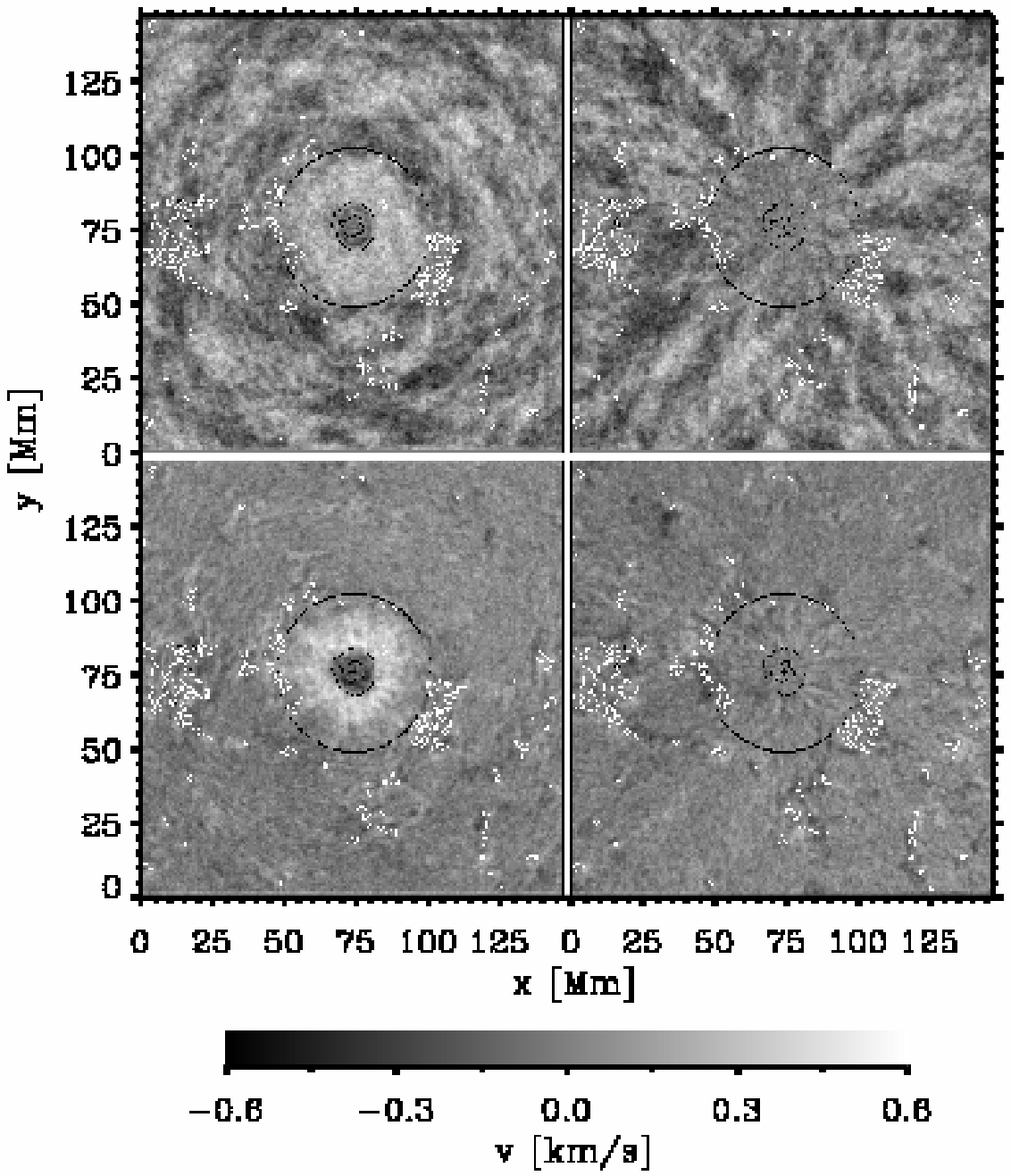}
   \caption{Radial (left) and tangential (right) horizontal velocities 
            for the WL series (top) and the UV series (bottom),  
            obtained from CLCT with a box size of four pixels.
            The dark lines mark the outer boundaries of the umbra and
            penumbra. The extension of the moat is given by the circle 
            with a radius of 27\,Mm. The white contours indicate the bright structures
            seen in the UV.
            The direction to disk center is downwards.
            Positive velocities are outward and counterclockwise, respectively.
               }
              \label{fig_lct4}
   \end{figure*}

    Waves propagating outward in areas outside the spots were detected by 
    Georgakilas et al. (\cite {georga2000}) from observations in the Fe~I~557.6~nm line.
    These waves have a phase velocity of about 0.5~km\,s$^{-1}$. Balthasar \&
    Schleicher (\cite{hb2008}) discovered Doppler-velocity features that survive for 
    45~minutes or even longer, traveling throughout the moat. These features 
    are restricted to the moat, not being detectable outside the moat.

    In this paper, we investigate a time series of a small sunspot and its 
    surroundings observed in two wavelength regimes with the spacecraft TRACE. 
    White-light images provide insight into the photosphere, and images at 170\,nm
    correspond to the layer just below the temperature minimum. 
    The magnetic properties of the same sunspot were determined by
    Balthasar \& G\"om\"ory (\cite{bago08}, hereafter Paper I). 
    This is an important difference from the investigations of 
    Sobotka \& Roudier (\cite{sobotka}) who applied a similar method as ourselves,
    but had no highly resolved magnetic data of their spots.


\section{Observations and data reduction}

We investigate two simultaneous time series obtained with TRACE on
May 27, 2006. For a detailed description of the instrument, we refer to
Handy et al. (\cite{TRACE}).
One series is in WL and the other in ultraviolet  
at 170\,nm (UV). The WL filter covers a wide wavelength range     
and reveals the layer at which the quiet sun's continuum at 500\,nm         
originates. According to Vernazza et al. (\cite{valc}), the UV         
continuum at 170~nm is formed 450 -- 500\,km higher in the quiet sun,  
just below the temparature minimum.
Uitenbroek (\cite{uit04}) computed the optical depth unity for         
the 170~nm continuum across a cross-section in a 3d snapshot of a simulation     
of solar convection.
The height of optical depth $\tau = 1$
varies between about 150 and 350\,km
and is, on average, located around 250\,km.

The field-of-view was centered on a small sunspot, NOAA 10886,
which was passing the central meridian at a heliographic latitude
of $7^\circ$~N. This spot had a radius of 6.7~Mm.
The series started at 00:39:30 UT, and had a cadence of 1~minute
and a total duration of 6~hours and 41~minutes. Moats can persist for 
this length of time, e.g.,
Sobotka \& Roudier (\cite{sobotka}) found moats that are stable     
over a period of 12\,h.

Data reduction largely followed that described in detail in Muglach
(\cite {karin03}). We subtracted a background and applied a flatfield
to correct for the loss of sensitivity of the central part of the CCD.
The image sequences were compensated for differential solar rotation and
the orbital image drift was taken care of using cross-correlation
methods. To remove intensity fluctuations caused by p--mode
oscillations, we also applied a $k{-}\omega$ filter to the WL  
series with a cutoff phase velocity of $v_{ph}$ = 3 km\,${\rm s^{-1}}$  .          

Figure \ref{fig-images} shows a selected area containing the spot.
The size of the area is 400 $\times$ 400 pixel, and the
size of a pixel is 0\farcs 5 corresponding to 370\,km.
Outside the spot, the WL-images are dominated by the granular structure, which
serves as a tracer for LCT. The single UV-images show many bright points moving 
radially with respect to the center of the spot, indicated by the elongated features 
in the averaged image.

To determine horizontal motions, we employ two different 
LCT techniques.
Here we compare the
results of two different LCT codes. The first is a classical 
LCT code (CLCT) that is the Lockheed Martin's Solar and 
Astrophysical Laboratory's version based on the original works of 
November \& Simon (\cite{lct}).
This method was used before e.g., ~by Deng et al. (\cite{nadeng06},
\cite{nadeng07}). The second method is 
Fourier local correlation tracking (FLCT) presented by
Fisher \& Welsch (\cite{fiwe08}).
An extensive comparison of these two and several other LCT methods based   
on synthetic magnetograms is presented by Welsch et al.~(\cite{2007ApJ...670.1434W}).
With their selection of parameters, they achieved a slightly superior performance
for FLCT compared to CLCT (see their Fig.\,6).

We correlate images with a time difference of eight minutes thus producing 401
velocity maps, which we average to obtain the final velocity map.
Both LCT methods need a spatial parameter as input, which preselects 
the spatial scales on which the horizontal velocities are determined.
For CLCT, one has to select the width
of the correlation box. We selected the values for $x$ and $y$ identically
and tried different values. The value of four pixels corresponding to 1.47~Mm
seems to be the most suitable selection for us because this value fits the size of   
granules in WL and that of the bright features in UV.   
A Gaussian bell, which has a width of 1.5 times the box size, is applied to the 
boxes as a weighting function. Shifts greater than 60\% of the box size,
but not smaller than three pixels, are clipped.
FLCT also uses a Gaussian windowing function. Each pixel is
multiplied by a Gaussian of width $\sigma$, where $\sigma$ is chosen by
the user. The final box size is determined by a cut-off value of the image,
which is set to be 0.1. In this way, the final box size is 
a factor of $\sim3$ greater than the
width of the Gaussian. A detailed description of the computational
approach including examples and runtime information of the FLCT is
given by Fisher \& Welsch (\cite{fiwe08}). The software is public and
can be downloaded \footnote{
http://solarmuri.ssl.berkeley.edu/overview/publicdownloads/software. html}.

Both methods give estimates of the horizontal motions in both the $x$ and $y$ direction.
These values are transformed into both a radial and tangential component
using the equations
   $$ v_{\rm r} = v_{\rm x}\cos\varphi + v_{\rm y}\sin\varphi \quad {\rm and} $$
   $$ v_{\rm t} = - v_{\rm x}\sin\varphi + v_{\rm y}\cos\varphi ,$$
where $\varphi$ is the azimuth angle measured about the center of the spot.

\section{Results}

   \begin{figure}
   \includegraphics[width=8.6cm]{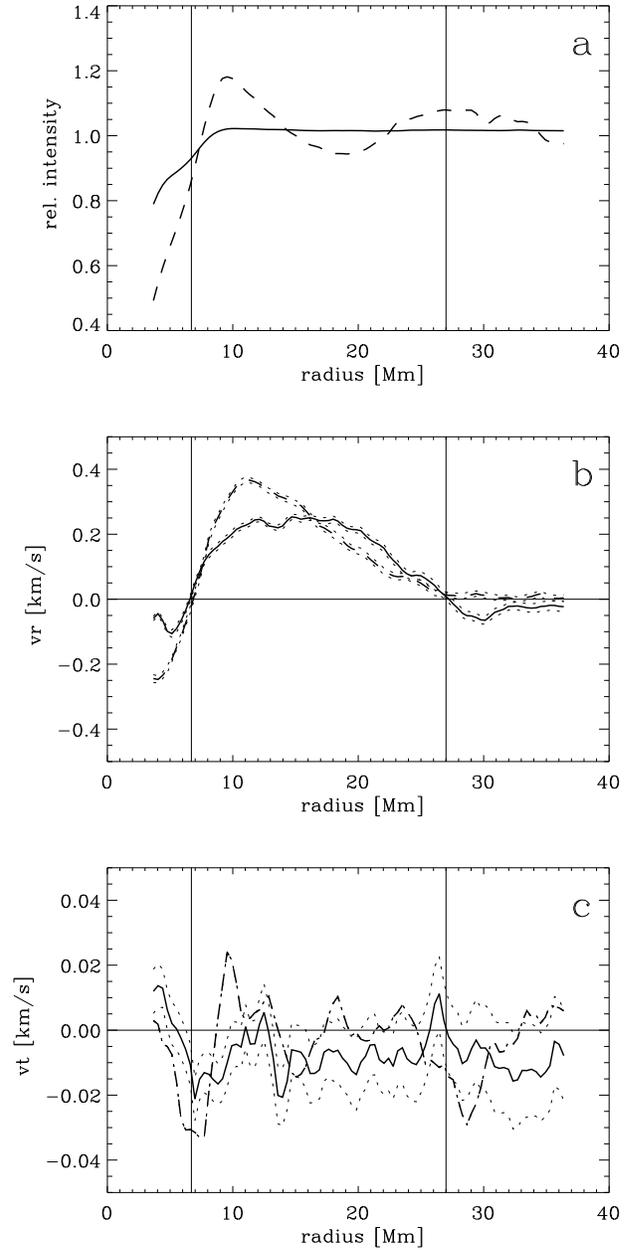}
      \caption{Radial dependence of relative intensities (a), radial velocity 
            component (b), and tangential velocity component (c), 
            averaged over the azimuth angle $\varphi$. WL data are given
            by the solid lines, UV data by dashed ones.
            Error ranges are indicated by the dotted lines. 
            Vertical lines mark the outer boundaries of penumbra and moat.
            The zero position of the radius corresponds to the center 
            of the spot marked in Fig~\ref{fig-images}. 
            The box size used in this figure is 4. Note the different scales for
            radial and tangential velocity components! 
            }
         \label{fig_xvonr}
   \end{figure}

   \begin{figure*}
   \centering
   \includegraphics[width=12cm]{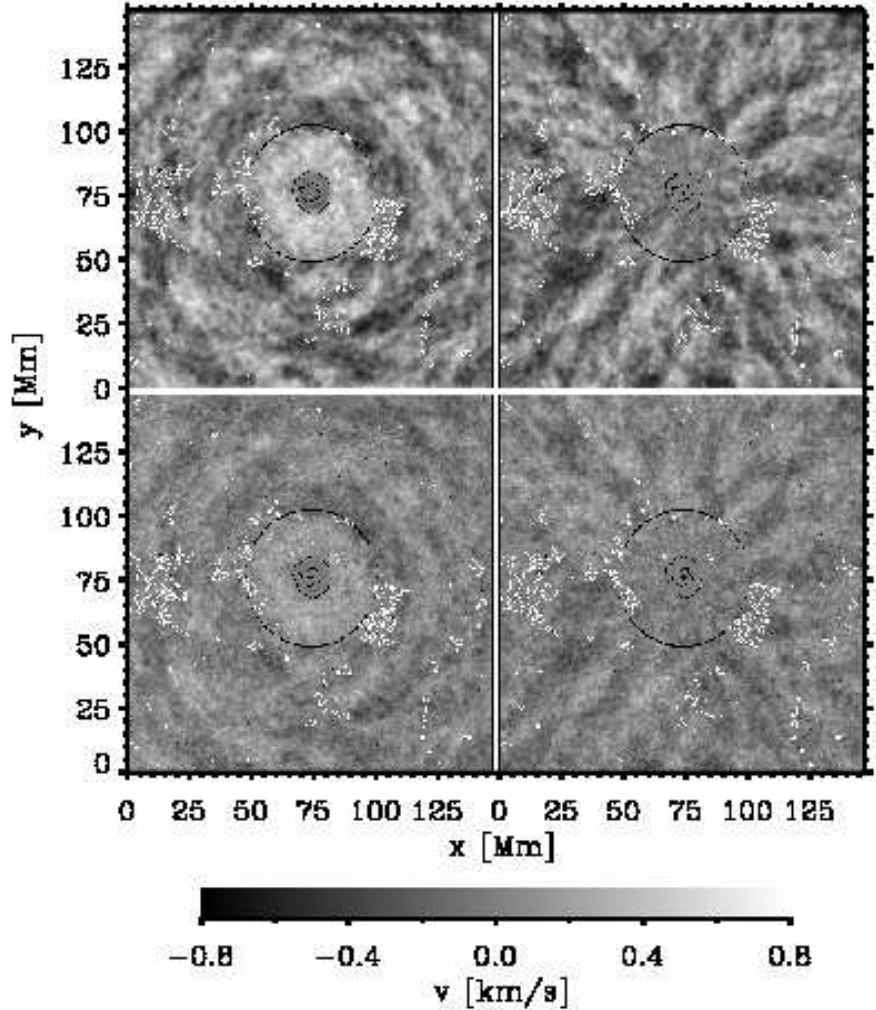}
   \caption{Radial (left) and tangential (right) horizontal velocities 
            for the WL series from CLCT (top) and  
            from FLCT (bottom),
            obtained with a box size parameter of 12 pixels for CLCT
            and $\sigma = 4$\,pixels for FLCT.
               }
              \label{fig_kmhbwl}
   \end{figure*}

\subsection {Comparison of white light and UV series}   

Horizontal flows were determined independently for the WL and the
UV series.
Maps of the radial and tangential velocities derived by CLCT
are shown in Fig. \ref{fig_lct4} for a box width of four pixels.
We detect radial outflows everywhere in the moat for both wavelength ranges. 
Velocities are higher in the UV for the inner moat, while WL exhibits higher
velocities in the outer moat.
For the inner moat, our detection of higher velocities in UV agrees
with the results of Sobotka \& Roudier (\cite{sobotka}).
Most of the structures that we track in WL are granules with
a short lifetime. Thus, it is not surprising that we see small-scale structures
in the WL-velocities. The bright points in the UV images survive for longer, and
we find more elongated structures in the UV-velocities. Tangential flows 
are rather small in the moat. There is a weak tendency for a negative value
in WL, which would imply with a clockwise motion. The radial dependence of the 
velocities after averaging over the azimuth with respect to the center of the 
spot is depicted in Fig. \ref{fig_xvonr}. The outflow of the moat ends at 27~Mm
for both WL and UV images.
The moat radius of 27\,Mm is roughly four times the spot radius of 6.7\,Mm.
Here, we observe a quite extended moat compared to the statistical values
published by Brickhouse \& LaBonte (\cite{brick}) and Sobotka \& Roudier
(\cite{sobotka}).
The extent of the flow coincides quite well with the distance to plage
areas and network points next to the spot as seen in the UV image.
These structures can also be recognized in the magnetic maps in Paper I.
Therefore, we consider this distance to represent the outer boundary of 
the moat.

In the penumbra, we see a few locations with inward motions.
Obviously this is not caused by penumbral grains that are known to move
outward in the outer penumbra. Their size is probably too small (below the 
TRACE resolution) to affect our results.

Beyond 27~Mm, the WL curve shows 
a flow towards the spot, while the UV curve remains close to zero.
The moat is surrounded by supergranules that manifest themselves 
as horizontal outflows from their respective centers.
Therefore, it is to be expected that the moat is surrounded
by motions toward the spot, as we see in WL. At a distance of roughly 50~Mm,
on the far side of the supergranules, we detect another ring of outflow.
These alternations repeated several times with increasing distance from the spot,
but the relation to the spot becomes less pronounced.
No systematic flow can be detected in UV close to the
moat. The supergranular flow is no longer present in those layers 
in which the UV light originates.
The supergranular flows also dominate the maps of tangential WL velocities    
outside the moat. Supergranules alternate between positive and negative 
velocities in the tangential direction, and this pattern is reversed in the next ring 
of supergranules. 
For this spot, we do not detect significant asymmetry in the east-west
direction as found for other ones by Sobotka \& Roudier (\cite{sobotka}).

\subsection {Comparison of the two methods}

   \begin{figure*}
   \centering
   \includegraphics[width=12cm]{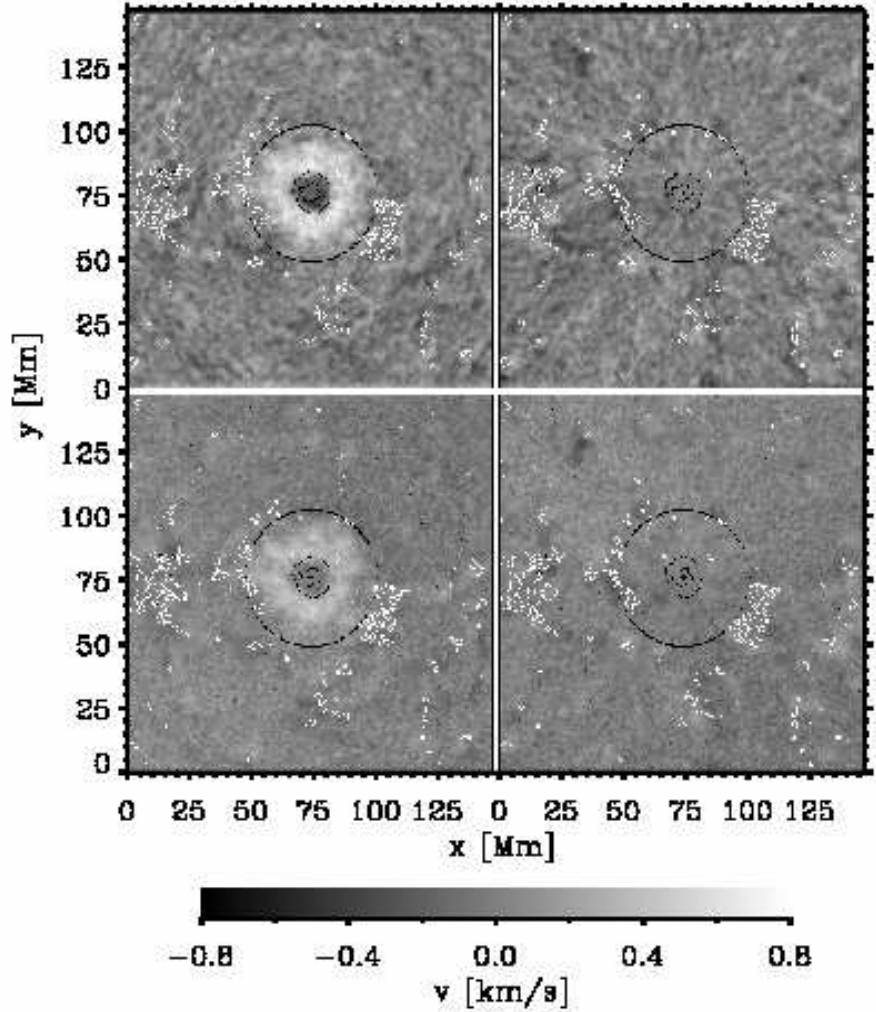}
   \caption{Same as Fig.~\ref{fig_kmhbwl} but for UV series.
                          }
              \label{fig_kmhbuv}
   \end{figure*}

To compare the two methods, the parameters must be selected in such a way
that differences caused by this selection are minimized.
Using CLCT, one chooses the box size as the main parameter, and
for FLCT the main parameter is the width $\sigma$ of the Gaussian bell.
The corresponding selection for
CLCT is a box size of 12 pixels and a factor of 0.33 for the Gaussian bell.
In Figs. \ref{fig_kmhbwl} and \ref{fig_kmhbuv}, we compare the results 
of the two methods. The large-scale structures 
in the velocity field are similar for both methods, but
CLCT velocities cover a wider range. The spatial resolution is 
of course lower than for a box size of four pixels as e.g., in Fig.~\ref{fig_lct4},
but surprisingly we measure higher velocities for the larger box size.
For both methods, we see the outflow that covers the entire moat.
Beyond the moat, we detect in WL the ring-like structures of positive
and negative radial velocity components and the alternation in the
signs of the velocity for the
tangential component, which are signatures of the supergranular velocity
field.

Outside the moat, results from CLCT applied to the UV data exhibit 
a patchy structure that resembles that of solar oscillations with typical 
lengths of about 5\arcsec. Superimposed is a faint indication of ring segments in 
the radial component and a radial structure in the tangential one, but the width of 
these structures is smaller than in the WL data.

\begin{figure}
   \includegraphics[width=8.6cm]{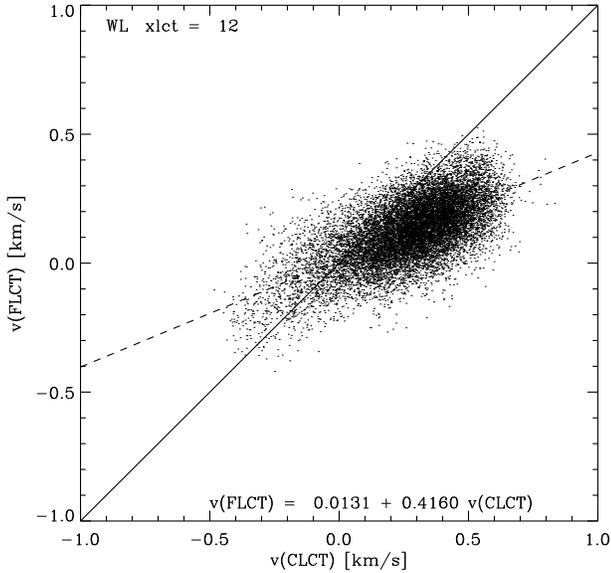}
      \caption{Comparison of the derived radial velocities from FLCT versus
            CLCT, restricted to the moat area. The dashed line represents the
            linear regression fit given in the lower part of the figure.
            The solid line is the 1:1 diagonal. The box size (CLCT) is 12 pixels
            and the width of the Gaussian bell, $\sigma$, (FLCT) is four pixels.
                        }
         \label{fig_vgl_10wl}
   \end{figure}

\begin{figure}
   \includegraphics[width=8.6cm]{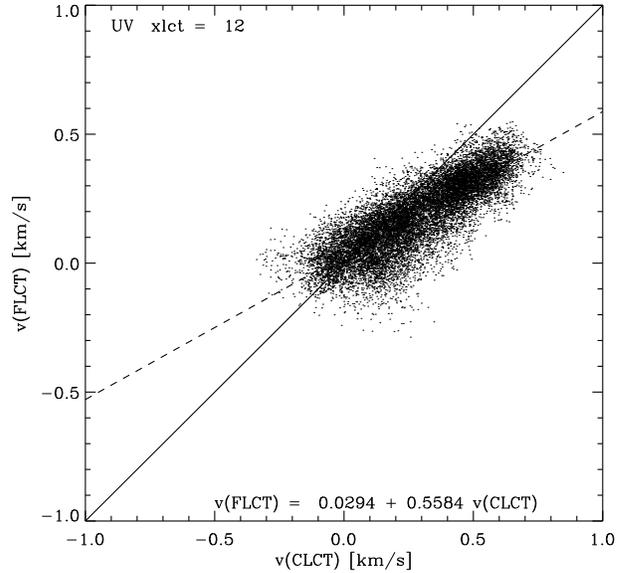}
      \caption{Same as Fig. \ref{fig_vgl_10wl} but for UV.
                        }
         \label{fig_vgl_10uv}
   \end{figure}

To present the difference more quantitatively, we calculated linear
regression fits of the radial velocities from FLCT compared to those from CLCT.
Here, we concentrate on the moat and therefore suppress all values both inside the spot
and beyond the mean moat radius of 27\,Mm. Coefficients of these fits 
are given in Table~\ref{tabcoef}. Scatter diagrams for the box size of 12 pixels
are shown in Figs.\,\ref{fig_vgl_10wl} and \ref{fig_vgl_10uv}.
Regression coefficients are smaller than one, indicating that CLCT velocities are
generally higher than FLCT velocities, but the scatter shows that 
a linear regression is a good approximation.

It is our impression that CLCT is more suitable for the presently investigated
case of the moat flow, where small individual structures such as UV bright points  
and MMFs move through their surroundings, because smaller correlation areas can be 
selected.
 
\begin{table}                                               
\caption{Linear regression fits$^a$ for the radial velocities in the moat
from FLCT versus those from CLCT.
}
\label{tabcoef}
\centering
\begin{tabular}{lrr|cc}
       $\lambda\lambda$  & box    & $\sigma$ & a$_0$&a$_1$   \\
                         & (CLCT) & (FLCT)   &      &        \\
\hline 
 & & & \\
WL &  4 &  4 & 0.036 $\pm$ 0.003 & 0.563 $\pm$ 0.014 \\
UV &  4 &  4 & 0.054 $\pm$ 0.004 & 0.781 $\pm$ 0.018 \\
WL & 10 & 10 & 0.041 $\pm$ 0.008 & 0.819 $\pm$ 0.019 \\
UV & 10 & 10 & 0.093 $\pm$ 0.007 & 0.678 $\pm$ 0.018 \\
WL & 15 & 15 & 0.030 $\pm$ 0.008 & 0.784 $\pm$ 0.018 \\
UV & 15 & 15 & 0.083 $\pm$ 0.007 & 0.685 $\pm$ 0.019 \\
WL & 12 &  4 & 0.013 $\pm$ 0.003 & 0.416 $\pm$ 0.010 \\
UV & 12 &  4 & 0.029 $\pm$ 0.004 & 0.558 $\pm$ 0.012 \\

\end{tabular}
\begin{list}{}{}
\item[$^{\mathrm{a}}$] Given are the coefficients from equation
                       $v_{\rm{r}}(\rm{FLCT})=a_0 + a_1 v_{\rm{r}}(\rm{CLCT})$
\end{list}
\end{table}

\section{Discussion}

In the inner moat, the proper motions measured in the UV are dominated by the bright  
points, which probably correspond to MMFs. MMFs exhibit high velocities of up to 
2\,km\,${\rm s^{-1}}$  (see Hagenaar \& Shine \cite{Hagenaar}), which are higher than velocities     
obtained from the granular pattern (0.3 -- 0.5\,km\,${\rm s^{-1}}$; see Sobotka \& Roudier 
\cite{sobotka} and Vargas Dom\'\i nguez et al.
\cite{vargas08}) or from spectroscopy (Balthasar et al. \cite {FPI96}).
The UV-bright points are much rarer in the outer moat. Thus, the faint granular
pattern affects the LCT results more than in the inner moat.
The extension of the moat of this spot is quite large, i.e., four times the
radius of the spot. Statistically, on average, only a factor of 1 or 2 is
expected, as found by Brickhouse \& Lites (\cite{brick}) and Sobotka \& Roudier
(\cite{sobotka}).
Although this spot was in its decaying phase, its area remained almost 
constant for several days.
Therefore, we cannot interprete the extension of the moat as a relict of an   
earlier larger spot. It is more likely to be related to the magnetic configuration, 
since the moat  
flow fills the space given by the magnetic network. Whether
this extended moat affects the decay of the spot remains speculative. 


Originally, MMFs were assumed to be pieces of the sunspot
that have been detached from the main spot and are being
swept away by the radial outflow of the convection around
the spot (as in supergranules, e.g.,~Simon \& Leighton, 1964).
In their seminal paper, Harvey \& Harvey (1973) find
that MMFs transport net flux away at the same rate as the flux
decay of the spot. They, therefore, consider MMFs as a signature
of sunspot decay (Kubo et al.~2007, 2008).

On the other hand, several studies have proposed
that (at least some) MMFs are an extension
of both the penumbra and the Evershed flow
(Zhang et al.~2003; Sainz Dalda \& Mart{\'\i}nez Pillet~2005;
Cabrera Solana et al.~2006; Vargas Dom\'\i nguez et al.~2007).
According to this interpretation, MMFs remain magnetically
connected to the central sunspot.

MMFs are identified with their magnetic signature.
In Paper I, the magnetic properties of this active region
were determined from full Stokes polarimetry. 
These data were taken on the same day, but after the TRACE sequences.
Therefore, we cannot use them to identify MMFs directly
by their polarization signature. In addition, we do
not have high resolution MDI magnetograms in hand
for the time of the TRACE observations. Nevertheless, we
can use the bright points in UV as proxies of the MMFs
(compare the averaged UV images and MDI magnetograms
in Muglach, 2003). Thus, the flow velocities in the UV
represent the velocities of the magnetic structures in the moat.
Magnetic field strengths derived from full Stokes polarimetry
of MMFs are mostly well below 1000\,G (Kubo et al.~2007).
The mechanism of flux expulsion can accumulate a field to the limit
approximately given by the equipartition of the magnetic
energy density of flux tubes and the kinetic energy density of
the granular flows (Parker 1963).
This equipartition field strength is around $\sim$ 500 G,
similar to the field strengths of MMFs.
Hence, the MMFs cannot significantly
alter the granular flow field. We conclude that
the flow velocities in WL represent the outward 
flow of the unmagnetized plasma, as the intensity 
contrast in WL is mainly caused by the granular pattern.

In Fig.\,4b, we see that the radial velocities in WL and
UV are different and this difference is larger than the
error in the measurement. In the UV, the apparent movement of the
bright points dominate the LCT results. This suggests that the MMFs
are not simply swept along the moat outflow in a completely
passive way, but
any detailed explanation of the physical processes causing
the faster motion of the MMFs would be
rather speculative at the present time.

\section{Conclusions}

Our most important results are summarized as follows:
   \begin{enumerate}
      \item The moat is filled with a horizontal outflow detectable in WL as well as in UV.
            In the inner moat, we detected higher velocities in UV than in WL and the 
            converse for the outer moat. 
            This outflow ends for both wavelength ranges at about 27~Mm
            (four times the radius of the spot) away from the center of the spot. 
      \item In the penumbra, we detected inward motions, mainly in the UV. However,
            the TRACE pixel size of 0\farcs5 is too coarse to perform a 
            sophisticated investigation of penumbral fine structures.
      \item Outside the moat, the velocity pattern of the supergranulation 
            was observed
            in WL as ring structures in the radial component and as 
            alternate positive and negative azimuthal values in the tangential component.
      \item CLCT was found to infer larger velocities than FLCT, and FLCT was found to 
            probably have limitations to its applicability if features moving at 
            different velocities occur close to each other.
   \end{enumerate}

Higher spatial resolution is needed in future investigations, and
strictly simultaneous spectropolarimetric observations would 
help to clarify many open questions such as
the relation between moat flow and Evershed effect, or to what extent the 
UV bright points are identical to MMFs.
For the visible range, 
these observations will become possible with the next generation of ground-based telescopes
(e.g., GREGOR, Volkmer et al., \cite{GREGOR}), which are currently under construction.

\begin{acknowledgements}
     We acknowledge the open data policy of the TRACE team. 
     We also thank G.~Fisher and B.~Welsch for help with the FLCT code
     and A. Antunes for running it on the NRL parallel computer cluster. 
\end{acknowledgements}

\end{document}